\documentclass[12pt]{article}
\usepackage{a4wide}
\usepackage{epsfig}
\usepackage{amsmath}

\newlength{\absize}
\catcode`@=11
\def\citer{\@ifnextchar
[{\@tempswatrue\@citexr}{\@tempswafalse\@citexr[]}}

\def\@citexr[#1]#2{\if@filesw\immediate
  \write\@auxout{\string\citation{#2}}\fi
  \def\@citea{}\@cite{\@for\@citeb:=#2\do
    {\@citea\def\@citea{--\penalty\@m}\@ifundefined
       {b@\@citeb}{{\bf ?}\@warning
       {Citation `\@citeb' on page \thepage \space undefined}}%
\hbox{\csname b@\@citeb\endcsname}}}{#1}}
\catcode`@=12

\begin{document}
  \thispagestyle{empty}
  \pagestyle{empty}
  \renewcommand{\thefootnote}{\fnsymbol{footnote}}
\newpage\normalsize
    \pagestyle{plain}
    \setlength{\baselineskip}{4ex}\par
    \setcounter{footnote}{0}
    \renewcommand{\thefootnote}{\arabic{footnote}}
\newcommand{\preprint}[1]{%
  \begin{flushright}
    \setlength{\baselineskip}{3ex} #1
  \end{flushright}}
\renewcommand{\title}[1]{%
  \begin{center}
    \LARGE #1
  \end{center}\par}
\renewcommand{\author}[1]{%
  \vspace{2ex}
  {\Large
   \begin{center}
     \setlength{\baselineskip}{3ex} #1 \par
   \end{center}}}
\renewcommand{\thanks}[1]{\footnote{#1}}
\begin{flushright}
\end{flushright}
\vskip 0.5cm

\begin{center}
{\large \bf $q$-Deformed Dynamics and Virial Theorem}
\end{center}
\vspace{1cm}
\begin{center}
Jian-zu Zhang$^{a,b, \S}$
\end{center}
\vspace{1cm}
\begin{center}
$^a$  Institute for Theoretical Physics, Box 316,
East China University of Science and Technology,
Shanghai 200237, P. R. China \\
$^b$ Department of Physics,
University of Kaiserslautern, PO Box 3049, D-67653  Kaiserslautern,
Germany 
\end{center}
\vspace{1cm}

\begin{abstract}
In the framework of the $q$-deformed Heisenberg algebra the investigation 
of $q$-deformation of Virial theorem explores that $q$-deformed quantum
mechanics possesses better dynamical property.
It is clarified that in  the  case of
the zero potential  the theoretical  framework for the $q$-deformed Virial
theorem is  self-consistent.
In the selfadjoint states the $q$-deformed uncertainty relation essentially 
deviates from the Heisenberg one.

\end{abstract}

\begin{flushleft}
${^\S}$ E-mail address:  jzzhang@physik.uni-kl.de  \\
\hspace{3.1cm} jzzhangw@online.sh.cn
\end{flushleft}
\clearpage
In hunting new physics at extremely high energy scale clarifying 
modification of the ordinary quantum mechanics at extremely small 
space scale is important.
The ordinary quantum mechanics is based on the Heisenberg 
commutation relation.
From the space scale $10^{-8}$ cm down to, according to the 
present test of quantum electrodynamics, at least $10^{-18}$ cm 
every test confirms that the Heisenberg commutation relation is
correct.
There is a possibility that the Heisenberg commutation relation
at extremely short distances much smaller than $10^{-18}$ cm
may need to be modified. 
In search for such possibility $q$-deformed quantum mechanics
is a candidate.
In literature different frameworks of $q$-deformed quantum mechanics were
established \citer{Schwenk,JZZ01}.
The framework of the $q$-deformed Heisenberg algebra developed
in Refs.~\cite{Hebecker,Fichtmuller} shows
interesting physical content and dynamical properties.
Its relation to the corresponding $q$-deformed boson commutation 
relations and the limiting process of $q$-deformed harmonic 
oscillator to the undeformed one are clear. 
The $q$-deformed uncertainty relation undercuts 
the  Heisenberg minimal uncertainty relation  
\cite{JZZ98,JZZ99,OZ00}.
 The non-perturbation energy spectrum of the $q$-deformed Schr\"odinger
 equation exhibits an exponential structure \cite{LW,Fichtmuller,JZZ00},
corresponding to new degrees of freedom and new quantum numbers
\cite{JZZ00}.
The perturbation expansion of the $q$-deformed Hamiltonian
possesses a complex structure, which amounts to some additional
momentum-dependent interaction
\cite{Hebecker,Fichtmuller,JZZ00,ZO01,JZZ01}.
A reliable foundation for the perturbation calculations 
in  $q$-deformed dynamics is established \cite{ZO01,JZZ01}.

In this paper we study the $q$-deformation of Virial theorem 
to demonstrate that $q$-deformed quantum
mechanics possesses better dynamical property.
In the ordinary quantum mechanics there is a delicate point in the
theoretical treatment of Virial theorem for the case of the zero
potential.
We clarified that in  the  case of the zero potential  the theoretical  
framework for  $q$-deformed Virial theorem is  self-consistent.
The demonstration of such self-consistency of $q$-deformed Virial theorem
is equivalent to the problem of finding a selfadjoint extension for the
representation of the $q$-deformed Heisenberg algebra.
Furthermore, we find that in the selfadjoint states the $q$-deformed 
uncertainty relation essentially deviates from the Heisenberg one.

In terms of $q$-deformed phase space variables - the position operator $X$
and the momentum operator $P$, the following $q$-deformed Heisenberg
algebra has been developed \cite{Hebecker, Fichtmuller}:
\begin{equation}
\label{Eq:q-algebra}
q^{1/2}XP-q^{-1/2}PX=iU, \qquad
UX=q^{-1}XU, \qquad
UP=qPU,
\end{equation}
where $X$ and $P$ are hermitian and $U$ is unitary:
$X^{\dagger}=X$, $P^{\dagger}=P$, $U^{\dagger}=U^{-1}$.
Comparing with the Heisenberg algebra, the operator $U$ is a new member,
called the scaling operator.
The necessity of introducing the operator $U$ is as follows.

The simultaneous hermitian of  $X$ and  $P$ is a delicate point in
 $q$-deformed dynamics.
The definition of the algebra
(\ref{Eq:q-algebra}) is based on the definition of the hermitian momentum
operator $P$.
However, if $X$ is assumed to be a hermitian operator in a Hilbert space,
the  $q$-deformed derivative \cite{Wess}
\begin{equation}
\label{Eq: q-derivative}
\partial_X X=1+qX\partial_X,
\end{equation}
which codes the non-commutativity of space, shows that
 the usual quantization rule $P\to -i\partial_X$ does not yield a
hermitian momentum operator. 
A hermitian momentum operator $P$ is related to
$\partial_X$ and $X$ in a nonlinear way by introducing a scaling operator
$U$ \cite{Fichtmuller}
\begin{equation}
\label{Eq:scaling}
U^{-1}\equiv q^{1/2}[1+(q-1)X\partial_X], \qquad
\bar\partial_X\equiv -q^{-1/2}U\partial_X, \qquad
P\equiv -\frac{i}{2}(\partial_X-\bar\partial_X),
\end{equation}
where $\bar\partial_X$ is the conjugate of $\partial_X$.
In Eq.~(\ref{Eq:q-algebra}) the parameter $q$ is a fixed real number.
It is important to make distinctions for different realizations of
the $q$-algebra by different ranges of $q$ values \citer{Zachos,Solomon}.
Following Refs.~\cite{Hebecker,Fichtmuller}  we only consider the case
$q>1$
 in this paper. The reason is that such choice of  the parameter $q$
leads to consistent dynamics.
In the limit $q\to 1^+$ the scaling operator $U$ reduces
to the unit operator, thus the algebra (\ref{Eq:q-algebra}) reduces to
the Heisenberg commutation relation.

The scaling operator $U$  plays every essential roles in $q$-deformed
 quantum mechanics.
The nontrivial property of  $U$  implies that the algebra
 (\ref{Eq:q-algebra}) has a richer structure than the Heisenberg
 commutation relation.
The operator $U$ is introduced in the definition of the hermitian
momentum,
thus it closely relates to properties of dynamics.
 $q$-deformed effects of dynamics are  determined by the  operator $U.$
Furthermore,  the  operator $U$ is crucial for guaranteeing the
self-consistency of the theoretical  framework in $q$-deformed  dynamics.

{\bf (I) $q$-deformation of Virial theorem}
In order to discuss Virial theorem in $q$-deformed dynamics, we need to
derive
the commutation relation between the momentum and the potential.
In the following we separately consider two cases:
 the regular potential which is singular free,
$V(X)= \sum_{n=0}^\infty \frac{V^{(n)}(0)}{n!}X^n$ and
singular  potential  $V(X)=V_0 X^{-n}, (n=1, 2, 3, \cdots).$

 From the  algebra (\ref{Eq:q-algebra})  the following commutation
relation
between $X$ and $P$ are obtained
\begin{equation}
\label{Eq:[X,P]}
[X,P]=iG, \quad G=\frac{U+U^{\dagger}}{q^{1/2}+q^{-1/2}}.
\end{equation}
Then from Eq.~(\ref{Eq:[X,P]}), carefully consider the ordering of the
 non-commutative operators $X$ and $U$, we obtain
\begin{equation}
\label{Eq:[P,Xn]}
[P,X^n]=-i\sum_{k=0}^{n-1} X^kGX^{n-1-k}
=\frac{-i}{q^{1/2}+q^{-1/2}}\frac{q^n-1}{q-1}
\left(U+q^{-(n-1)}U^{\dagger} \right)X^{n-1},
\end{equation}
where $(n=0, 1, 2, 3, \cdots).$

The derivation of the commutator  $[P,X^{-n}]$
proceeds as follows.
Time Eq.~(\ref{Eq:[X,P]}) by $X^{-1}$ from the left,  take  conjugation of
the obtained equation,
then time the conjugated equation by $X^{-1}$ from the left again, we
obtain
\begin{equation}
\label{Eq:[X,P]1}
[X^{-1},P]=-iX^{-1}GX^{-1}.
\end{equation}
From Eq.~(\ref{Eq:[X,P]1}),  carefully consider the ordering of the
 non-commutative operators $X^{-1}$ and $G$, we obtain
\begin{eqnarray}
\label{Eq:[P,X-n]}
[P,X^{-n}]&=&-i\sum_{k=1}^n  X^{-k}GX^{-(n+1)+k}
 \nonumber \\
&=&\frac{i}{q^{1/2}+q^{-1/2}}\frac{q(q^n-1)}{q-1}
\left(q^{-(n+1)}U+U^{\dagger} \right) X^{-(n+1)},
\end{eqnarray}
where $(n=1, 2, 3, \cdots).$ In the above in the last step the
relation between $X^{-1}$ and $U,$  which are obtained from
the algebra (\ref{Eq:q-algebra}), is used,
\begin{equation}
\label{Eq:U-X-1}
UX^{-1}=qX^{-1}U. \end{equation}

In Eq.~(\ref{Eq:[P,Xn]}) let $n=-m,$ then
 Eq.~(\ref{Eq:[P,Xn]}) is turned into  Eq.~(\ref{Eq:[P,X-n]}).
We conclude that  Eq.~(\ref{Eq:[P,Xn]}) works for
$n=0, \pm1, \pm2, \pm3, \cdots$.
Thus we define  the  $q$-deformed derivative $\mathcal{D}_X X^n$  as
\begin{equation}
\label{Eq:q-deriv}
\mathcal{D}_X X^n \equiv i [P,X^n]
=\frac{1}{q^{1/2}+q^{-1/2}}\frac{q^n-1}{q-1}
\left(U+q^{-(n-1)}U^{\dagger} \right) X^{n-1}, 
\end{equation}
where $(n=0, \pm1, \pm2, \pm3, \cdots)$.
In the limit $q\to 1^+$   Eq.~(\ref{Eq:q-deriv}) reduces to the ordinary
result: $\mathcal{D}_X X^n \to \; nX^{n-1}.$
From Eq.~(\ref{Eq:q-deriv}) it follows that
the commutation relation between the momentum and the potential.reads
\begin{equation}
\label{Eq:[P,V(X)]}
[P,V(X)]=-i\mathcal{D}_X V(X),
\end{equation}
where $\mathcal{D}_X V(X)\equiv \sum_{n=1}^\infty \frac{V^{(n)}(0)}{n!}
\mathcal{D}_X X^n$  (for regular potential),
 $\mathcal{D}_X V(X)\equiv V_0 \mathcal{D}_X X^{-n}$  (for singular
potential), 
and $\mathcal{D}_X X^n$ is defined by Eq.~(\ref{Eq:q-deriv}).

Now it is ready to derive   the $q$-deformed Virial theorem.
The $q$-deformed phase space ($X$, $P$) governed by the $q$-algebra
(\ref{Eq:q-algebra}) is a $q$-deformation of 
phase space ($\hat x$, $\hat p$) of the ordinary quantum mechanics
 governed by the Heisenberg commutation relation $[\hat x, \hat p]=i,$
 thus all machinery of  the ordinary quantum mechanics can be applied to
 the $q$-deformed quantum mechanics.
It means that dynamical equations of a quantum system are the same
for the undeformed phase space variables  ($\hat x$, $\hat p$)  and
for the $q$-deformed  phase space variables ($X$, $P$), that is,
 the $q$-deformed Schr\"odinger equation with potential $V(X)$ is
\begin{equation}
\nonumber
i \partial_t |\psi(t)\rangle=H|\psi(t)\rangle,  \quad
H(X,P)=\frac{1}{2\mu}P^{2}+V(X).
\end{equation}

Now we consider the time derivative of the average of the operator 
$XP,$ 
\begin{equation}
\label{Eq:[XP,H]}
i\frac{d}{dt} \overline {XP}=\overline {[XP,H]},
\end{equation}
where $\overline {F}\equiv\langle \psi(t)|F|\psi(t)\rangle.$
From Eqs.~(\ref{Eq:q-algebra}) and  (\ref{Eq:[X,P]}) we obtain
\begin{equation}
\nonumber
[XP,P^2]=iPGP+iGP^2=i(q^{1/2}+q^{-1/2})^{-1}
\left[(1+q^{-1})U+(1+q)U^{\dagger}\right]P^2.
\end{equation}
For a  stationary state $|\psi(t)\rangle$ we have
 $\frac{d}{dt} \overline {XP}=0.$
Using the above results from  Eq.~(\ref{Eq:[XP,H]})
 it follows that  the $q$-deformed Virial theorem reads
\begin{equation}
\label{Eq:Virial}
\frac{1}{q^{1/2}+q^{-1/2}}
\overline {\left[(1+q^{-1})U+(1+q)U^{\dagger}\right]T}
=\overline {X\mathcal{D}_X V(X)},
\end{equation}
where the undeformed kinetic energy $T=P^2/(2\mu).$ 
When $q\to1^+,$   Eq.~(\ref{Eq:Virial}) reduces to the Virial theorem
in the ordinary quantum mechanics:
 $2\overline {T}=\overline {X V'(X)}.$

The outstanding characteristic in the formulation of  the $q$-deformed
Virial theorem is the factor of the scaling operator $U$.
Generally in $q$-deformed quantum mechanics the  dynamical formulas
are much complex than ones in the ordinary quantum mechanics
because of such $q$-deformed factors involved in the formulation.
The $q$-deformed  Virial theorem shows a complicated structure:
in Eq.~(\ref{Eq:Virial}) except the complex $q$-deformed derivative 
on the right hand side,
 there are the factors of the scaling operator $U$ on the left hand set.
In the following we demonstrate that just because of such $q$-deformed
factors
 the  dynamical behavior of the $q$-deformed  Virial theorem is improved
for
the zero potential case.

{\bf (II)  Self-consistency in the zero potential case.}
In the ordinary quantum mechanics Virial theorem does not work
 for the case $V(X)=0.$
\footnote{In the ordinary quantum mechanics in order to avoid
Eq.~(\ref{Eq:[XP,H]}) for a stationary state leading to  an
 inconsistent result in  the zero potential case,
 instead of taking  the sate  $|\psi(t)\rangle$ as plane wave state,
$\langle x|\psi_p(t) \rangle\sim exp[i(px-Et)],$
  normalizable wave packet state
\begin{eqnarray*}
\nonumber
&&\langle x|\psi_{wp}(t) \rangle \sim
\int_{-\infty}^{\infty}[cosha(k-k_0)]^{-1} exp[i(kx-\omega(k)t)]dk.  \\
&&\omega(k)=\omega(k_0)+\left(\frac{d\omega}{dk}\right)_{k=k_0}(k-k_0)
\end{eqnarray*}
should be used.}
But the $q$-deformed Virial theorem  is  self-consistent in this case.
The demonstration of such self-consistency of
$q$-deformed Virial theorem is equivalent to the problem of finding a
correct selfadjoint extension for the representation of the $q$-deformed
Heisenberg algebra (\ref{Eq:q-algebra}).

In the following we first review the results about selfadjoint extension
of
the algebra (\ref{Eq:q-algebra}) obtained in Refs.
\cite{Hebecker,Wess99},
 then demonstrate the self-consistency of $q$-deformed Virial theorem
in the zero potential case.
Following \cite{Wess99} we start from the Hilbert space
$\mathcal{H}^{\sigma}_s.$ Its basis is defined as
\begin{equation}
\label{Eq:Hs}
X|n,\sigma \rangle^s=\sigma sq^n|n,\sigma \rangle^s, \quad
\sigma=\pm1,\quad  1<s<q.
\end{equation}
For fixed $\sigma$ and $s,$ from the algebra  (\ref{Eq:q-algebra}) it
follows that 
$$XU|n,\sigma \rangle^s=\sigma sq^{n+1}U|n,\sigma \rangle^s, \;
XU^{\dagger}|n,\sigma \rangle^s=\sigma sq^{n-1}U^{\dagger}|n,\sigma \rangle^s,$$
thus up to a phase factor we have
\begin{equation}
\label{Eq:Hs1}
U|n,\sigma \rangle^s=|n+1,\sigma \rangle^s, \;
U^{\dagger}|n,\sigma \rangle^s=|n-1,\sigma \rangle^s.
\end{equation}
However, on the basis $|n,\sigma \rangle^s$  the momentum $P$ is not a
selfadjoint linear operator, but has selfadjoint extension. Define states
\begin{equation}
\label{Eq:HsI}
|\tau p_{\nu},\sigma \rangle^s_I=\frac{1}{\sqrt{n}}N_q
\sum_{n=-\infty}^\infty q^{n+\nu}\{cos_q(q^{2(n+\nu)})|2n,\sigma \rangle^s
+i\tau sin_q(q^{2(n+\nu)})|2n+1,\sigma \rangle^s.
\end{equation}
\begin{equation}
\label{Eq:HsII}
|\tau p_{\nu},\sigma \rangle^s_{II}
=U^{\dagger}|\tau p_{\nu},\sigma \rangle^s_I.
\end{equation}
where $\tau=\pm1;$ $0\le\nu<\infty;$  
$cos_q(x)$ and  $sin_q(x)$ are $q$-deformed cosine and
sine functions \cite{KS}.
The states  (\ref{Eq:HsI}) and (\ref{Eq:HsII}) are eigenstates of $P:$
\begin{equation}
\label{Eq:P-HsI-II}
P|\tau p_{\nu},\sigma \rangle^s_I
=\frac{1}{s\lambda q^{1/2}}\sigma \tau q^{2\nu}
|\tau p_{\nu},\sigma \rangle^s_I, \quad
P|\tau p_{\nu},\sigma \rangle^s_{II}
=\frac{1}{s\lambda q^{1/2}}\sigma \tau q^{2\nu-1}
|\tau p_{\nu},\sigma \rangle^s_{II},
\end{equation}
where $\lambda=q-q^{-1}.$
 The states  (\ref{Eq:HsI}) and (\ref{Eq:HsII}) form an orthogonal 
and complete set in  the Hilbert  space $\mathcal{H}^{\sigma}_s,$
thus define a selfadjoint extension of $P.$
But this representation is reducible.
In 
order to obtain irreducible representation, according to
$\sigma=+$ and  $\sigma=-,$ rearrange  the Hilbert space as
$\mathcal{H}^{+}_s\oplus\mathcal{H}^{-}_s,$ and
on this sum Hilbert space  define states
\begin{eqnarray}
\label{Eq:irr}
&&|\tau p_{\nu}\rangle^s_{I}=\frac{1}{\sqrt{2}}\{ |\tau
p_{\nu},+\rangle^s_{I}
+ |-\tau p_{\nu},-\rangle^s_{I}\}, \nonumber\\
&&|\tau p_{\nu}\rangle^s_{II}=\frac{1}{\sqrt{2}}\{ |\tau
p_{\nu},+\rangle^s_{II}
+ |-\tau p_{\nu},-\rangle^s_{II}\}.
\end{eqnarray}

We discuss the property of the state (\ref{Eq:irr}): \\
(i) the  The state $|\tau p_{\nu}\rangle^s_{\alpha}$ is eigenstate 
of $P,$
\begin{equation}
\label{Eq:P-HsI}
P|\tau p_{\nu} \rangle^s_{\alpha}=
P_{\alpha}|\tau p_{\nu} \rangle^s_{\alpha}, \quad
(\alpha=I,II)
\end{equation}
with $P_I=\frac{1}{s\lambda q^{1/2}} \tau q^{2\nu}, \;
P_{II}=\frac{1}{s\lambda q^{1/2}} \tau q^{2\nu-1}.$ \\
(ii) On the selfadjoint extension state 
$|\tau p_{\nu}\rangle^s_{\alpha}$ the scaling 
operator $U$ is unitary.
From Eqs.~(\ref{Eq:Hs1})-
 (\ref{Eq:HsII}) and (\ref{Eq:irr}), we have.
\begin{equation}
\label{Eq:U-HsI}
U^{\dagger}|\tau p_{\nu}\rangle^s_I=|\tau p_{\nu}\rangle^s_{II}, \quad
U^{\dagger}|\tau p_{\nu}\rangle^s_{II}=|\tau p_{\nu-1}\rangle^s_I,
\end{equation}
thus
\begin{equation}
\nonumber
UU^{\dagger}|\tau p_{\nu}\rangle^s_{\alpha}=U^{\dagger}U|\tau
p_{\nu}\rangle^s_{\alpha}
=|\tau p_{\nu}\rangle^s_{\alpha}, \quad
(\alpha=I, II).
\end{equation}
(iii) The state (\ref{Eq:irr}) forms an orthogonal and complete set  
in the sum Hilbert space
$\mathcal{H}^{+}_s\oplus\mathcal{H}^{-}_s,$
\begin{equation}
\label{Eq:ort}
{}_{\alpha}^s\langle \tau p_{\nu}|\tau' p_{\nu'} \rangle^s_{\beta}
=\delta_{\alpha\beta}\delta_{\tau\tau'}\delta(\nu-\nu'), \; 
(\alpha, \beta=I, II), 
\end{equation}
\begin{equation}
\label{Eq:comp}
\sum_{\tau=\pm1}\sum_{\alpha=I,II}\int_{-\infty}^{\infty}d P_{\alpha}
|\tau p_{\nu} \rangle^s_{\alpha}{}_{\alpha}^s\langle \tau p_{\nu}|
=I.
\end{equation}
 The state (\ref{Eq:irr}) is the irreducible representation of 
the momentum operator $P.$

We conclude that  in the sum Hilbert space 
$\mathcal{H}^{+}_s\oplus\mathcal{H}^{-}_s$
on  the basis  (\ref{Eq:irr}) the $X$ and $P$ are
selfadjoint linear operators and $U$ is unitary one, thus define
a selfadjoint extension for the representation of
the $q$-deformed Heisenberg algebra (\ref{Eq:q-algebra}).\\
(iv) In the state $|\tau p_{\nu}\rangle^s_{\alpha}$
 the $q$-deformed uncertainty relation undercuts the Heisenberg minimal one.
From the algebra (\ref{Eq:q-algebra}) it follows that
\begin{eqnarray}
\label{Eq:X}
&&X=-i\lambda^{-1}P^{-1}(q^{1/2}U^{\dagger}-q^{-1/2}U), \\
&&X^2=-\lambda^{-2}P^{-2}[q^2U^{\dagger 2}-(q+q^{-1})+q^{-2}U^2].
\end{eqnarray}
In the state $|\tau p_{\nu}\rangle^s_{\alpha}$ the variances
of $X$ and $P$ are
\begin{equation}
\label{Eq:variances}
(\Delta X)^2=\lambda^{-2}P^{-2}_{\alpha}(q+q^{-1})=finite, \quad
(\Delta P)^2=0.
\end{equation}
Thus $(\Delta X)(\Delta P)=0,$ which undercuts the Heisenberg minimal
uncertainty relation. This shows that 
$q$-deformed uncertainty relation essentially deviates from the 
Heisenberg one.
 
Now we demonstrate the self-consistency of $q$-deformed Virial theorem
 for the zero potential case.
In this case the stationary state  $|\psi(t)\rangle$ is 
\begin{equation}
\label{Eq:irrt}
|\tau p_{\nu},t\rangle^s_{\alpha}
= exp(-iE_{\alpha}t)|\tau p_{\nu}\rangle^s_{\alpha}, \qquad
E_{\alpha}=\frac{1}{2\mu}P_{\alpha}^2, \qquad
(\alpha=I, II)
\end{equation}
For the  case $V(X)=0$ the right hand side of  Eq.~(\ref{Eq:Virial})
equals zero.
For the stationary state (\ref{Eq:irrt}) the left hand side of  
Eq.~(\ref{Eq:Virial}) is proportional to
\begin{eqnarray}
&&{}_{\alpha}^s\langle \tau p_{\nu},t|\left[(1+q^{-1})U+(1+q)U^{\dagger}
\right]T|\tau p_{\nu},t \rangle^s_{\alpha} \nonumber \\
&&=E_{\alpha}
\bigl[(1+q^{-1}) {}^s_{\alpha}\langle \tau p_{\nu}|U
|\tau p_{\nu}\rangle^s_{\alpha} 
+  (1+q)\;{}^s_{\alpha}\langle \tau p_{\nu}|U^{\dagger}
|\tau p_{\nu}\rangle^s_{\alpha} \bigr].
\nonumber
\end{eqnarray}
From  Eq.~(\ref{Eq:U-HsI}) we have
${}^s_{\alpha}\langle \tau p_{\nu}|U|\tau p_{\nu}\rangle^s_{\alpha}
={}^s_{\alpha}\langle \tau p_{\nu}|U^{\dagger}|\tau
p_{\nu}\rangle^s_{\alpha}=0.$
Thus for any given $P_{\alpha}$ the left hand side of
Eq.~(\ref{Eq:Virial})
also equals zero.

Now we discuss the time average in an arbitrary state 
$|\psi(t)\rangle.$ 
Expand the state $|\psi(t)\rangle$ according to 
the state (\ref{Eq:irrt})
\begin{equation}
|\psi(t)\rangle=\sum_{\tau,\alpha}\int dP_{\alpha}
C_{\tau,\alpha}(P_{\alpha})
 exp(-iE_{\alpha}t)|\tau p_{\nu}\rangle^s_{\alpha}.
\nonumber
\end{equation}
Take the time average of the left hand side of 
Eq.~(\ref{Eq:Virial})
\begin{eqnarray}
&&\lim_{T \to \infty}\frac{1}{T}\int_{-T/2}^{T/2}dt
\langle \psi(t)|\left[(1+q^{-1})U+(1+q)U^{\dagger}
\right]T|\psi(t)\rangle 
\nonumber \\
&&=\lim_{T \to \infty}\frac{1}{T}\int_{-T/2}^{T/2}dt
\sum_{\tau,\alpha}\sum_{\tau',\alpha'}
\int dP_{\alpha}dP'_{\alpha'}C^{\ast}_{\tau,\alpha}
C_{\tau',\alpha'}exp(i(E_{\alpha}-E'_{\alpha'})t)
\nonumber \\
&&\bigl[(1+q^{-1}) {}^s_{\alpha}\langle \tau p_{\nu}|U
|\tau' p_{\nu'}\rangle^s_{\alpha'} 
+  (1+q)\;{}^s_{\alpha}\langle \tau p_{\nu}|U^{\dagger}
|\tau' p_{\nu'}\rangle^s_{\alpha'} \bigr],
\nonumber
\end{eqnarray}
where $E'_{\alpha'}=\frac{1}{2\mu}P_{\alpha'}^{'2}, \;
(\alpha'=I, II)$
with $P'_I=\frac{1}{s\lambda q^{1/2}}\tau' q^{2\nu'}$ and
$P'_{II}=\frac{1}{s\lambda q^{1/2}}\tau' q^{2\nu'-1}.$
When $E_{\alpha}=E'_{\alpha'}$  we have 
${}^s_{\alpha}\langle \tau p_{\nu}|U|\tau p_{\nu}\rangle^s_{\alpha}
={}^s_{\alpha}\langle \tau p_{\nu}|U^{\dagger}|\tau
p_{\nu}\rangle^s_{\alpha}=0$ again.
When $E_{\alpha}\ne E'_{\alpha'}$ the oscillator factor
$\lim_{T \to \infty}\frac{1}{T}\int_{-T/2}^{T/2}dt
exp(i(E_{\alpha}-E'_{\alpha'})t) \to 0.$
Thus the time average of the left hand side of 
Eq.~(\ref{Eq:Virial}) is zero.

From the above results we conclude that in the zero potential case
the $q$-deformed Virial theorem is  self-consistent.

In $q$-deformed quantum mechanics, unlike  the ordinary quantum
 mechanics, the $q$-deformed commutation relation closely influence
dynamics.
The reason is that  the new  element $U$ of  the $q$-deformed
Heisenberg algebra  is introduced in the definition of the hermitian
momentum, thus the structure of the algebra  closely relates to properties
of dynamics.
The exploration of self-consistency of  $q$-deformed Virial theorem in
 zero potential case shows that $q$-deformed quantum mechanics possesses
a better dynamical behavior.
$q$-deformed quantum theory might be a realistic physical theory at
extremely
short distances  much smaller than  $10^{-18}$~cm.
It is  hoped that studies of $q$-deformed dynamics at the level
of quantum mechanics will give some clue for the further studies at the
level
of  more realistic field theory.

\vspace{0.4cm}
  This work has been supported by the Deutsche Forschungsgemeinschaft
(Germany). The project started during the author's visit at
Max-Planck-Institut f\"ur Physik (Werner-Heisenberg-Institut).
 He would like to thank  H.J.W. M\"uller-Kirsten for stimulating
 discussions.
 His work has also been supported by the National Natural Science
Foundation of China under the grant number 10074014 and by the Shanghai
Education Development Foundation.


\end{document}